# Linear build-up of Fano resonance spectral profiles

P.A. Golovinski[1,2], A.V. Yakovets[1], V.A. Astapenko[1]

[1]Moscow Institute of Physics and Technology (State University)
[2]Voronezh State Technical University

The build-up dynamics of a continuous spectrum under the action of a weak laser field on a Fano resonance with the use of the pulses with the Lorentz spectrum and ultrashort pulses in the wavelet form is investigated. A dispersion-time excitation dependence of the Fano resonances in a He atom, in an InP impurity semiconductor, in longitudinal optical LO-phonons of a shallow donor exciton in pure ZnO crystals, and in metamaterials are calculated. The numerical simulation of the dynamics has shown time-dependent formation a Fano spectral profile in the systems of different physical natures under the action of ultrashort pulses with attosecond and femtosecond durations.

## 1. Introduction

Excitation of an energy level by a laser pulse is a classical method of laser spectroscopy [1, 2], and for long pulses with a great number of oscillations it has been investigated in detail [3, 4]. When few-cycle pulses are used, there occur specific changes in the dynamics of a two-level system that manifest themselves in violation monotonicity of the time dependence the transition probability in weak fields and in a possibility to control population with the use of the chirped pulses [5-7]. In many real physical problems, an excited level is not isolated, and other degrees of freedom in the system can be taken into account as interaction of the level with an energy band. This is how autoionization states [8, 9], excited states of an atomic nucleus against the background of a continuous spectrum [10], or states of an impurity in a solid interacting with a matrix spectrum [11] are arranged. Resonances occurring in excitation of such states have a characteristic asymmetric shape due to interference of a level and the band states. The theory of resonances with a discrete state in continuum was developed independently by Fano and Feshbach, and resonances themselves that are manifested in a great diversity of physical systems are generally referred to as Fano-Feshbach resonances [12].

The development of physics and technology of nanosized structures has revealed the presence of new various Fano-Feshbach resonances in these structures [13] that we hereafter, according to the established usage, just call Fano resonances. For the prospects of development nanooptical and nanoelectronic devices, of particular interest are resonances observed in interaction of states in quantum dots with acoustical [14, 15] and optical [16-18] phonons as well as with plasmons in metal nanoparticles and nanowires [19-28].

Excitation of a level against the background of a continuous spectrum under the action a weak monochromatic field is described by Fano profile for an absorption line. The action of an ultrashort laser pulse of low intensity on such a system changes considerably the shape of an absorption profile as a function of the pulse carrier frequency [29]. The description of this dynamic effect can be made for low-intensity fields within the framework of the non-stationary perturbation theory for an alternating electromagnetic field, while for strong fields a situation occurs that implies taking into account nonlinear effects. For autoionization states of atoms, there is a consistent theory that made it possible to describe the evolution of such states excited by attosecond pulses [30-36]. The main result in this direction was an understanding both of the intra-atomic component of the dynamics of a time-dependent dipole response [37, 38] and the temporal pattern formation of electrons in a continuous spectrum [39], which was clearly confirmed by experimental measurements. For electrons in a continuous spectrum, there turns out to be important the competition of processes of direct ionization and transition to a continuous spectrum through excitation of a discrete state, which interference varies in time. Most of earlier works were dedicated to isolated resonances, but the influence of overlapping the resonances on their shape [40, 41] and dynamics [42] was also considered. For quantum dots, the exciton generation is observed in the visible region of light, where studying Fano resonances becomes easier due to the availability of the advanced technique of manipulation the single-cycle laser pulses [43, 44].

On the whole, quantum interference shows itself as a resonance in the Fano optical absorption spectrum with a characteristic asymmetric shape. In close proximity to the main maximum, a deep minimum occurs at an energy corresponding to destructive interference the transition amplitudes to a discrete state and continuous spectrum states. An important basis of the Fano theory is the assumption a sufficiently high density of states that can be assigned to continuous spectrum. In this case, great diversity of physical systems is described in a unified way in the linear approximation and demonstrates similar features of behavior. We focus our attention on the time resolution of the process induced by pumping a number of systems since an understanding the quantum dynamics of complex state formation of Fano resonances is practically important for obtaining data about a preparation efficiency of such states with the use of short pulses depending on a chosen system parameters.

## 2. Dynamics of Fano resonance in a weak optical field

The principal peculiarity of the set problem is necessity of taking into account both interaction of a discrete state with a continuum and interaction of the whole system with a pulsed field equally. The search for solution can be carried out with the use of the expansion of the non-stationary wave function in terms of the basis of unperturbed stationary wave functions of the unperturbed problem

$$\psi_E(t) = a(t)\psi_0 + \int dE' b_{E'}(t) u_{E'} . \qquad (1)$$

Here $\psi_0$ is the wave function of a discrete state, $u_E$ are the unperturbed states of a continuum of states, $V_E$ is the matrix element of interaction of a discrete state with a continuum. The simplest way to solve the problem is to find the system response to the $\delta$- kick that is suitable for description of pulses with the duration that is significantly less than the time of state decay. In this assumption, the non-stationary and stationary perturbations can be taken in to account in two steps due to their separation in time. The role of a $\delta$-pulse comes to producing a state being an initial condition for the problem with a stationary Hamiltonian. The final solution looks like [34]

$$A_E(t) = -id_E \frac{\left(e^{-iEt}(q+\varepsilon) - (q-i)e^{-i\overline{E}_r t}\right)}{\varepsilon + i}, \qquad (2)$$

where $d_E$ is the matrix element of the dipole amplitude for transition in continuum, $\varepsilon = 2(E-E_r)/\Gamma$ is the relative resonance detuning, $\overline{E}_r = E_r - i\Gamma/2$ is the complex quasi-energy of a resonant state, $E_r = E_0 + \Delta$ is the resonance energy. The resonance energy shift $\Delta(E)$ is given by the principal integral value

$$\Delta(E) = P\int dE' \frac{|V_E|^2}{E - E'}, \qquad (4)$$

where $V_E$ is the matrix element of interaction a discrete state with a continuum.

The Fano parameter

$$q = \frac{d_0 + P\int dE' \dfrac{d_E V_E}{E - E'}}{\pi V_E d_E} \qquad (5)$$

($d_0$ is the dipole matrix element of transition to a discrete state) can be determined from the experiment. It is equal to the ratio of the transition amplitude to the resonant state with a continuum ingredient in the transition amplitude to the unperturbed continuum that can be considered to be constant within a resonance. The continuum population is calculated according to the formula

$$P(E,t) = |A_E(t)|^2. \qquad (6)$$

The population being formed after cessation of a $\delta$-pulse follows the time dependence

$$P(E,t) = \frac{|d_E|}{\varepsilon^2 + 1} \left| e^{-i\varepsilon\Gamma t/2}(q+\varepsilon) - (q-i)e^{-\Gamma t/2} \right|^2. \qquad (7)$$

A linear response to a pulse of an arbitrary shape [32, 33] can be found on the basis of the equation (2) with the use of the Duhamel's integral, which brings us to the equation

$$A_E(t) = -id_E \left[ g(E,t)[q+\varepsilon] - (q-i)g(\overline{E}_r, t) \right], \qquad (8)$$

where

$$g(E,t) = \int_{-\infty}^{t} dt' \, e^{iE(t'-t)} f(t'). \qquad (9)$$

Here $f(t)$ is the strength of the electric field of a laser wave as a time function. We consider the energy of an initial unperturbed state to be a reference point equal to zero. The first term in the square bracket of the equation (8) is responsible for formation a Fano profile in a long time limit, that is, when the observation time exceeds considerably the time of decay a discrete state. The second term in this bracket describes a damping transient process. After completion of the pulse action we have

$$g(E, t \to \infty) = 2\pi e^{-iEt} F(\omega), \qquad (10)$$

where $F(\omega)$ is the Fourier transform of the pulse.

Excitation of Fano resonance by pulses with a structure described by an envelope both as in the form of the sine squared as in the Gaussian form was considered earlier [32]. We make the main emphasis on a pulse with a Lorentz spectrum that looks like a damped oscillation, that is,

$$f(t) = f_0 e^{-\lambda t} \sin \omega t, t > 0. \qquad (11)$$

For such a pulse

$$g(E,t) = \frac{if_0}{2} \left[ e^{-\lambda t} \left( \frac{e^{i\omega t}}{E+\omega+i\lambda} - \frac{e^{-i\omega t}}{E-\omega+i\lambda} \right) + \frac{2\omega e^{-iEt}}{(E+i\lambda)^2 - \omega^2} \right]. \qquad (12)$$

After cessation of the pulse action and termination the transient processes ($\lambda t \gg 1, \Gamma t \gg 1$), this function tends to

$$g(E,t) = \frac{if_0 \omega e^{-iEt}}{(E+i\lambda)^2 - \omega^2}, \tag{13}$$

and

$$P(E,\infty) \approx \frac{|d_{E_r} f_0/2|^2}{(E-\omega)^2 + \lambda^2} \frac{(q+\varepsilon)^2}{\varepsilon^2 + 1}. \tag{14}$$

Thus the final form of the continuum population is described by the product of the Lorentz profile of a laser pulse spectrum and an asymmetric Fano profile. For a spectrally narrow laser pulse with $\lambda \ll \Gamma/2$, it is possible to measure a Fano profile by the continuum population, tuning the laser frequency. For a spectrally wide laser pulse with $\lambda \gg \Gamma/2$, the distribution modulation is small, and Fano profile can be measured directly as a final population of continuum states. Hence, for ultrashort pulse and a spectrally narrow Fano resonance, the details of a pulse structure become unessential, and a rectangular spectrum is sufficient for adequate representation of a laser field.

For single-cycle or few-cycle laser pulses, of interest is the analytical approximation in the form of wavelet-type field, when the condition $\int_{-\infty}^{\infty} dt' f(t') = 0$ is undoubtedly fulfilled. As basis functions of wavelets [45], we choose derivatives of the Gaussian functions

$$h_m(t) = (-1)^m \frac{d^m}{dt^m} e^{-t^2} = H_m(t) e^{-t^2}. \tag{15}$$

Here $H_m(t)$ are the Hermitian polynomials [46]. At $m = 1$, a WAVE-wavelet arises, and at $m = 2$ there is a MHAT-wavelet, or a "Mexican hat". In the frequency domain $\tilde{h}_m(\omega) = \frac{1}{2\sqrt{\pi}}(-i\omega)^m e^{-\omega^2/4}$. For a pulse in the form of a basis wavelet $f(t) = f_0 h_m(t/\tau)$ with a characteristic duration $\tau \sim 1/E_r$, the function $g(E,t)$ after completion of a pulse is

$$g(E, t \to \infty) = \sqrt{\pi} \tau f_0 (-iE\tau)^m e^{-iEt - (E\tau)^2/4}, \tag{16}$$

and the population of the continuous spectrum states looks like

$$P(E,\infty) = P_0(E_r, \tau) \frac{(q+\varepsilon)^2}{\varepsilon^2 + 1}, \tag{17}$$

where the factor

$$P_0(E_r, \tau) = \pi \left|\frac{d_{E_r} f_0}{E_r}\right|^2 (E_r \tau)^{2(m+1)} e^{-(E_r \tau)^2/2} \tag{18}$$

takes into account the spectral density of a ultrabroadband pulse in the vicinity of a resonance. From the equations (17), (18) it follows that the population of the continuous spectrum states after cessation the action of a broadband pulse weakly depends on the pulse parameters that influence only the total magnitude of the population. The comparison of efficiency for excitation by the "Mexican hat" pulse with $E_r\tau = 1$ and a pulse with $E_r\tau = 2$, which maximum of the spectral intensity coincides with a resonance, gives 0.07 of the value of the probability of the first version as compared with the second one.

### 3. Examples of Fano resonance formation

Let us consider the dynamics of the band of states population, when the distance between band levels is much less than the value of a level-band interaction. In this case the band levels form a quasi-continuum. As an example of such states are autoionization states in atoms. Let us describe the dynamics of a build-up the autoionization resonance 2s2p in a helium atom under the action of ultrashort pulses of attosecond duration. The total dynamics of autoionization consists of a short-time quantum system response on the pulse that at present remains uninvestigated and of the evolution of a state after the action of a pulse [47]. For short times, the response depends on the details of the broadband pulse structure, that is, the spectral modulation of the carrier phase at the width of a resonance. The energy of such a doubly-excited state in a He atom, counted off from the ground state, is 60.15 eV at a width of 37 meV. In the experiment, a laser pulse with a spectral width of 400 meV was used. The Fano parameter for this resonance is $q = -2.77$ [48]. Fig. 1 shows the time dependence of continuum excitation by a laser pulse. Since the spectrum width of the experimental laser pulse is much more than the width of a resonance, the continuum state population after the passage of the pulse tends to the dispersion curve of a Fano profile plotted as the inset. In Fig. 2, two shadow pattern projections of the excitation intensity caused by broadband and narrow-band pulses with the Lorentz spectral distribution are presented for comparison. The broadband pulse excites the asymmetric Fano resonance, which is represented in Fig. 1 as the response surface. The broadband pulse excites an asymmetric Fano resonance that is presented in Fig. 1 as a response surface. For a narrow-band laser pulse the population pattern is slightly asymmetric only at first, and then the symmetric Lorentz distribution is formed as it is prescribed by the equation (14).

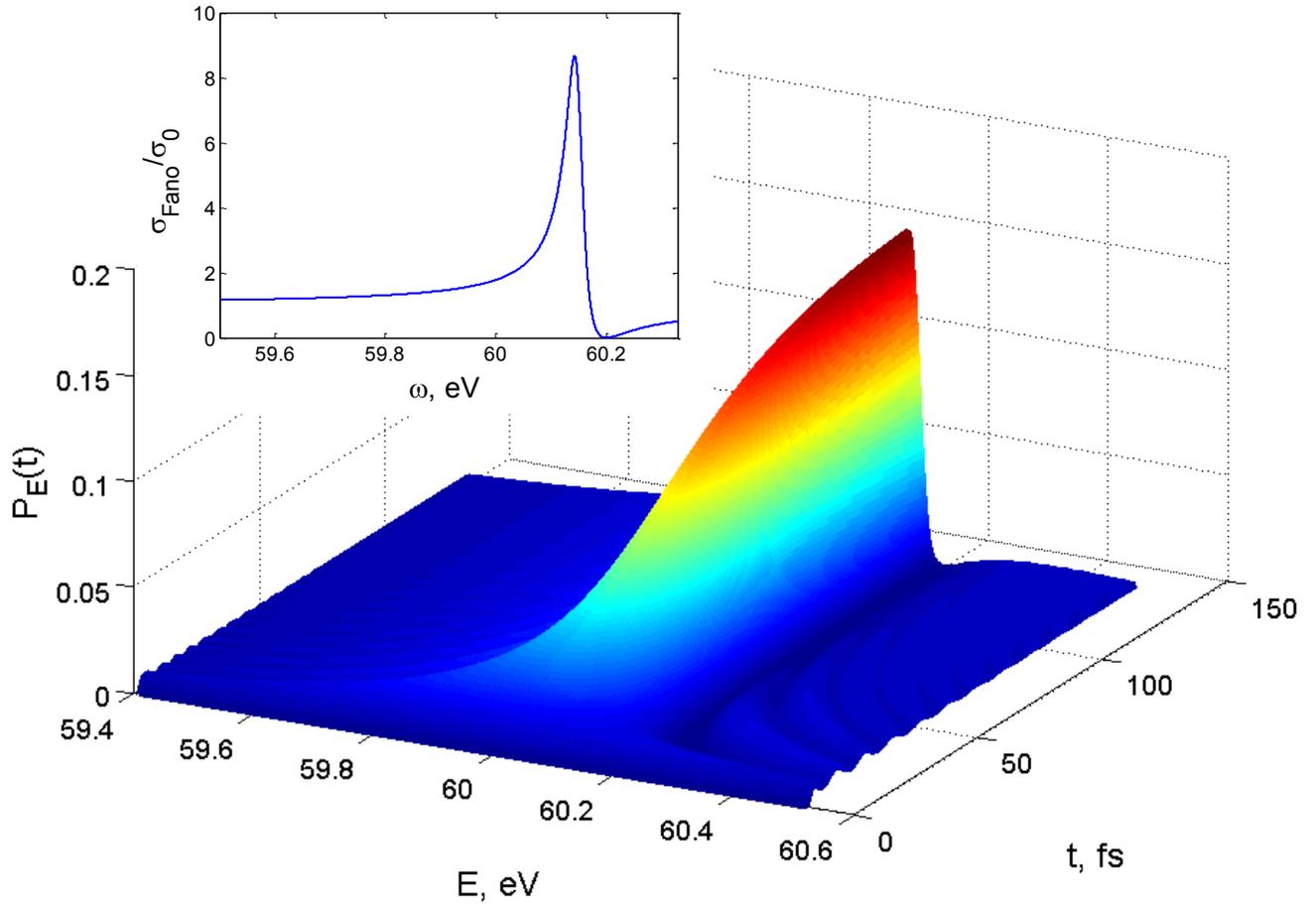

**Fig. 1.** Energy-time excitation dependence for Fano resonance in a He atom. The pulse field $f(t) = f_0 e^{-\lambda t} \sin \omega t$, $\lambda$ = 0.6 fs$^{-1}$. The carrier frequency coincides with the resonance. The Fano resonance dispersion in a He atom is presented as the inset.

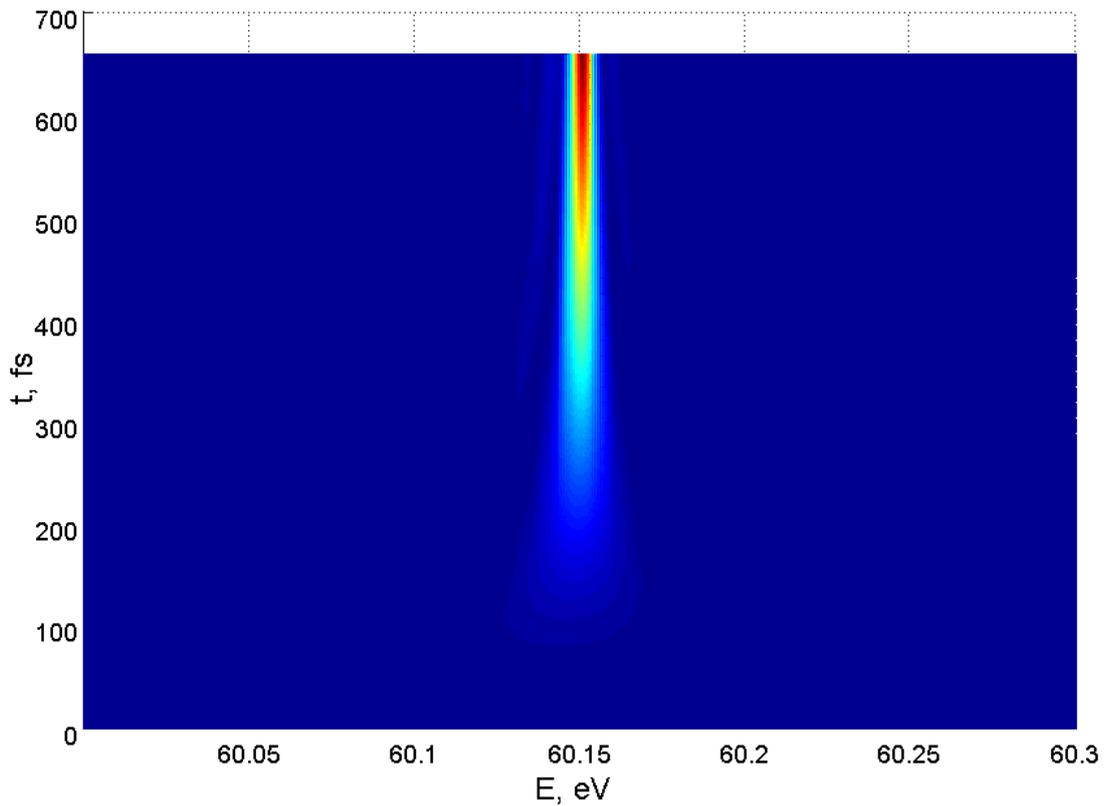

a)

b)

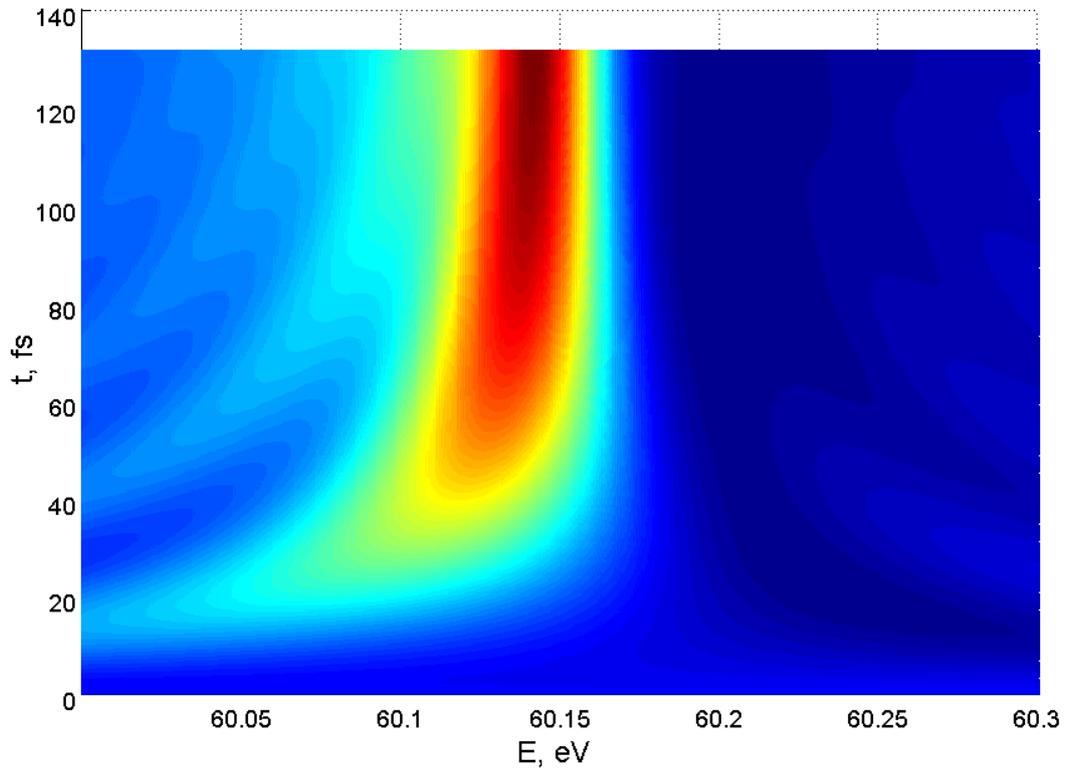

**Fig. 2.** The shadow patterns of the excitation intensity for broadband and narrow-band pulses with the Lorentz spectrum: a) $\lambda = 6$ fs$^{-1}$; b) $\lambda = 0.0015$ fs$^{-1}$.

Another example of Fano resonance gives us a photocurrent arising in some $A^{III}B^{V}$ and $A^{II}B^{VI}$ donor-type impurity semiconductors. In these semiconductors, shallow donor states are described by a hydrogen-like model [49]. The energy of resonant states lies in the interval of a continuum, when there are no phonons in the system. However, electrons interact with longitudinal optical phonons, which results in formation of Fano resonance. In InP samples, the resonance frequency is $\omega_0 = 42.85$ meV, the width of the resonance is $\Gamma = 0.84$ meV, and the Fano parameter is $q = 3.92$. These parameters make it possible to trace the build-up dynamics of Fano state under the action of ultrashort laser pulse, presented in Fig. 3.

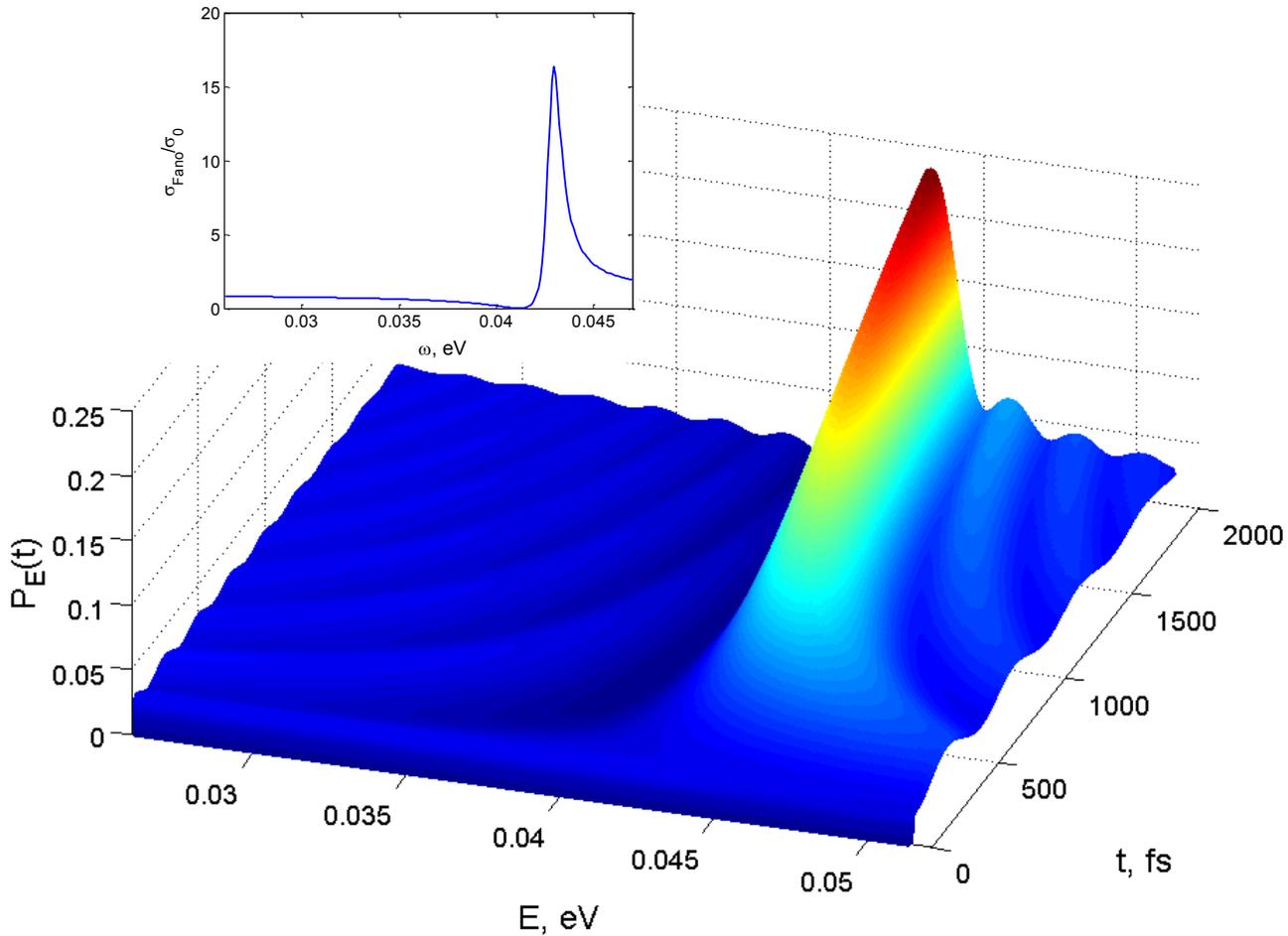

**Fig. 3.** Energy-time excitation dependence for Fano resonance in an InP impurity semiconductor. The pulse field $f(t) = f_0 e^{-\lambda t} \sin \omega t$, $\lambda$ = 0.015 fs$^{-1}$. The carrier frequency coincides with the resonance. The Fano resonance dispersion in an InP impurity semiconductor is presented as the inset.

There are rather few investigations of Fano resonance are known for the processes of solids photoemission. Despite of this, Fano resonance in phonon lines of longitudinal optical LO-phonons for photoluminescence of a shallow donor exciton in pure ZnO crystals was observed [50]. Here occurs resonant interaction of an electron with an optical phonon, and phonon mixing provides formation of a continuous spectrum that is necessary for Fano resonance. The strongest peak corresponding to a zero phonon line (that is, without absorption or emission of additional phonons) has a maximum at a frequency of an optical photon of 3.361 eV. The shape parameter $q$ for different lines is in a range of 1.43-1.85. The width of the resonance is ~2.5 meV at a temperature of 5 K. Fig. 4 presents the results of calculation of Fano resonance excitation in a donor exciton depending on the time. Complete overlapping the resonance interval by a laser pulse spectrum provides formation of a characteristic asymmetric distribution.

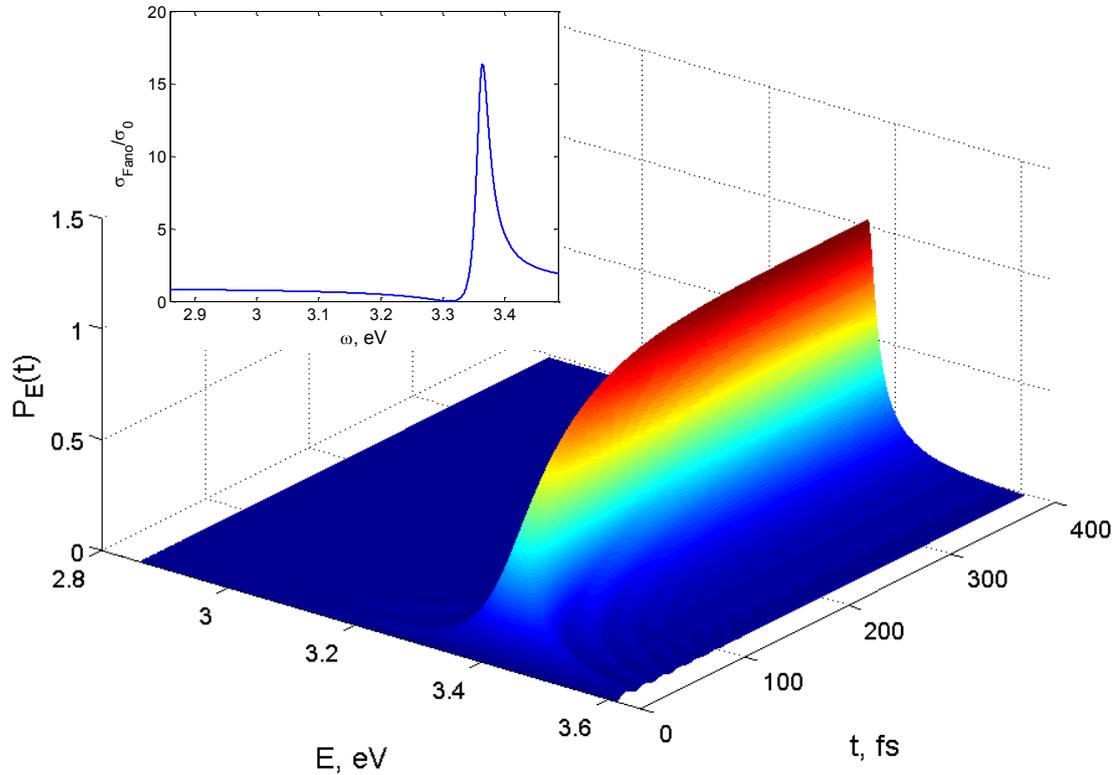

**Fig. 4.** Energy-time excitation dependence for Fano resonance phonon lines of the longitudinal optical LO-phonons for a shallow donor exciton in pure ZnO crystals excited by the field $f(t) = f_0 e^{-\lambda t} \sin \omega t$, $\lambda = 0.15$ fs$^{-1}$. The carrier frequency coincides with the resonance. The dispersion dependence is presented as the inset.

### 4. Asymmetric resonances in metamaterials

Fano resonances were found also in plasmon nanoparticles, photon crystals, and electromagnetic metamaterials [51-53], where it has a classical wave origin. Abrupt dispersion of Fano resonance profile is promising for use in sensors, laser generation, switches, nonlinear devices, and slow light systems. Fano resonances show themselves in reflection and transmission of periodic metal structures, that is, in metal photon crystals. Fano resonances are diverse; for example, an array of gold nanowires placed on a single-mode waveguide in the form of a plate has a Fano resonance manifesting itself in extinction of transverse electric (TE) polarization (an E-field parallel to gold wires) due to the connection between the lattice and the waveguide. The narrow mode of the waveguide interferes with a continuum of vacuum states. Light in the waveguide demonstrates destructive interference due to reradiated light by the particles plasmons. As a light angle or a lattice spacing change, spectral dispersion shows pronounced anticrossing that is an indicator of polaritons in the form of a waveguide-particle plasmon. In linear and nonlinear experiments, a Fano line can be well described by a model of two coupled classical oscillators with narrow and wide resonances [54]. It should be noted that the results obtained on the basis of the model of coupled classical oscillators in fact have a more wide range of applicability since linear response of a system is fully determined by its dispersion properties. This follows from a possibility to represent pulses of arbitrary shapes as the Fourier integral. Therefore the specific realization a response as linear operator with the empirical spectrum, whether it be a classical or quantum system, has no effect on the linear properties being described. This in turn makes it possible to apply our theoretical description of the quantum Fano resonance excitation by a weak field to the description of classical Fano resonance in metamaterials.

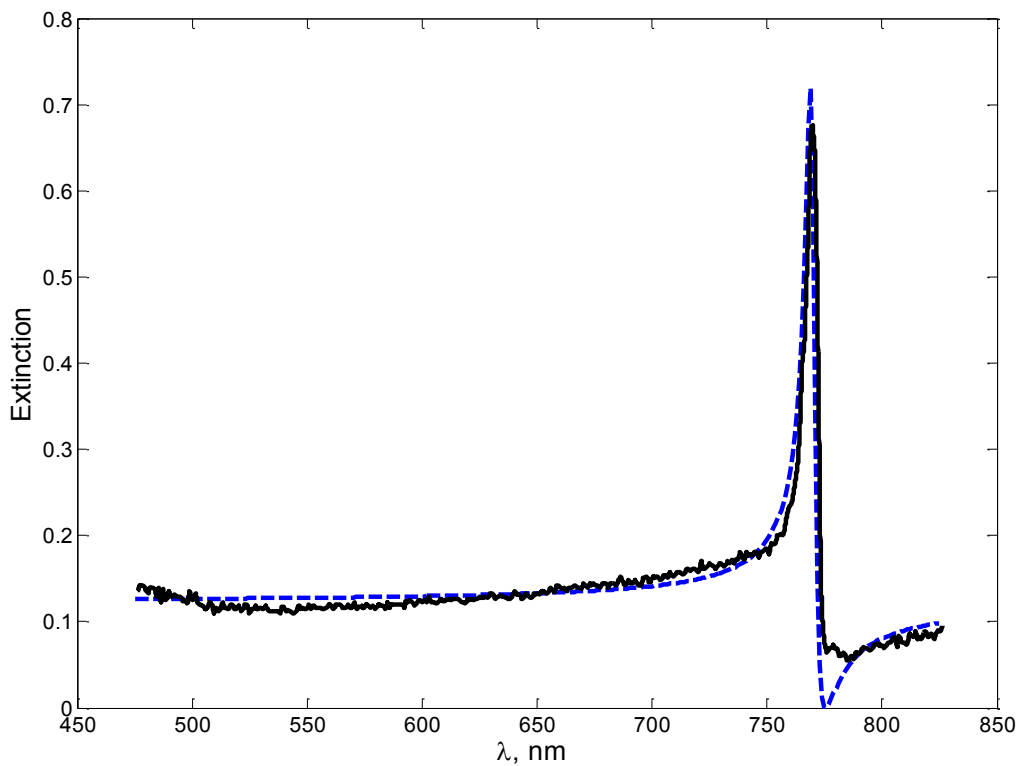

**Fig. 5.** Fano profile for the light extinction in a metal photon crystal. Solid curve shows experimental data [54], dotted curve represents theoretical data.

Fig. 5 shows an experimentally measured Fano resonance in a metal photon crystal [51] consisting of a gold nanowire lattice placed on a single-mode film waveguide of tin oxide, in which light falls normally to the structure, and profile analytical approximation. The analytical form of a Fano resonance profile $\sim (q+\varepsilon)^2/(1+\varepsilon^2)$ allows one to describe experimental curves by a proper choice of the parameter $q$. As a result of the approximation, the values of the parameters $\lambda_0 = 770$ nm for the frequency of a resonance, $\Delta\lambda = 5$ nm for its width, and the shape parameter $q = 2.2$ were obtained. The time response of a system can vary significantly depending on the time-shape of an acting pulse. We have carried out the calculation for a pulse, the spectral width of which is much more than the width of a resonance. Because in this case the details of the pulse shape are unessential for description the dynamic of resonance, we have replaced it by a $\delta$-pulse, for which a time-dependent linear response is described by the equation (10). Fig. 6 presents extinction time dependence of ultrabroadband spectrum when pulse passing through the structure. The first stage of excitation is short-term and is absent in the model. After formation the initial state, the process with duration $t \sim 1/\Gamma$ takes place, as a result of which there remains a stationary spectrum.

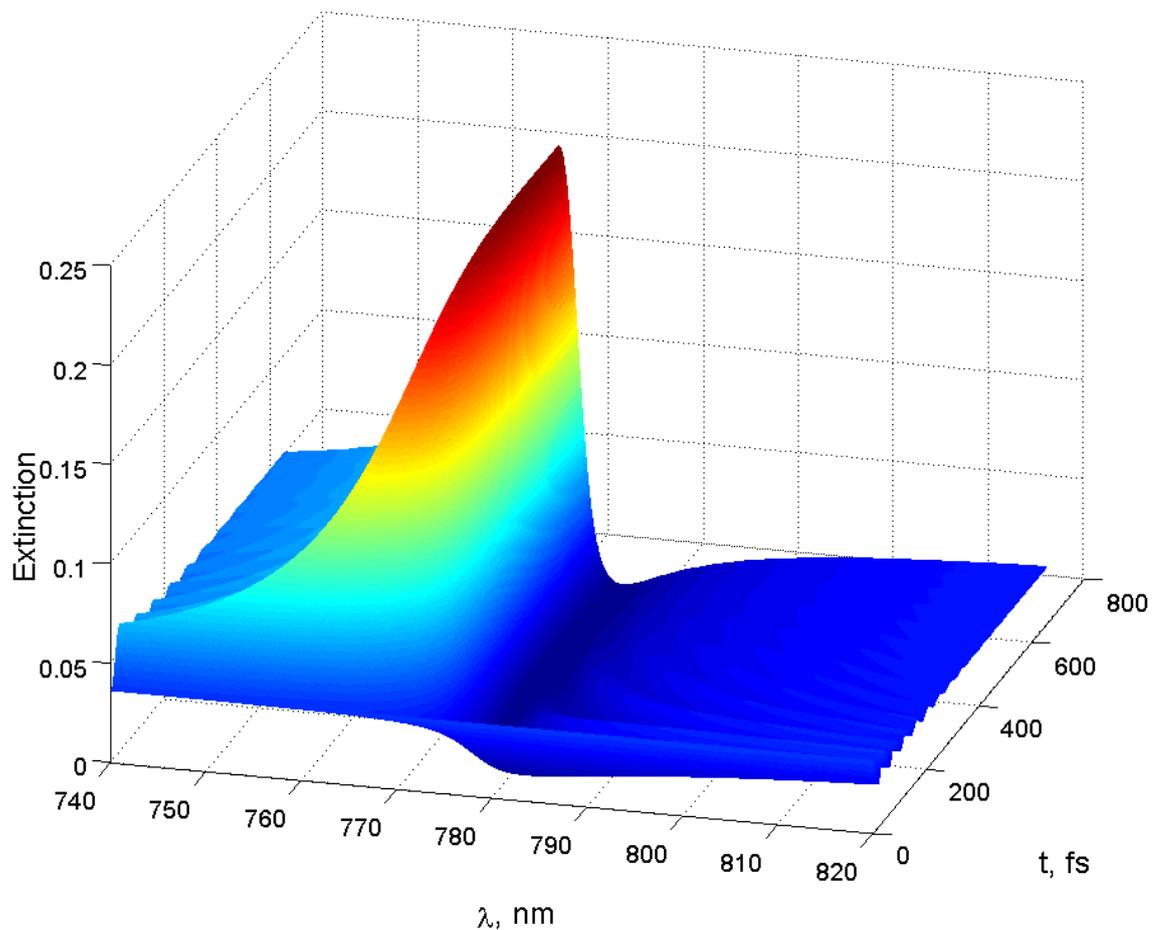

**Fig. 6.** The extinction dynamics of a broadband pulse by a photon crystal.

## 5. Conclusions

The analysis of the excitation dynamics of a continuous spectrum under the action of a weak laser field on Fano resonance by pulses, having a Lorentz spectrum, and ultrashort pulses in the wavelet form shows formation the population, containing two interfering amplitudes. The first amplitude depends on the energy of a continuous spectrum state being populated and on the time. It tends to a stationary distribution after finishing a laser pulse field action. The second amplitude does not depend on the given energy in spectrum and takes into account the relaxation of a discrete resonant state due to its decay to the continuum states. This term tends to zero with time. The rate of nulling of this correction is adjusted by the width of a quasi-stationary state. If the width of a pulse spectrum exceeds the width of a Fano resonance, the pulse action ends before termination of the decay of a discrete state to a continuum, and after cessation of the pulse action the transient process still goes on. If the width of a pulse spectrum is less than the width of Fano resonance, only part of the resonance band of a Fano quasi-continuum is excited, but by the time of cessation of the pulse action the processes of decay of a discrete state to a continuum are completely finished. The final population of a quasi-continuum is determined by the product of a Fano resonance shape and a pulse spectrum. So if a laser pulse is long, the excitation intensity reproduces the shape of a pulse spectrum, but with amplitude specified by a Fano profile at the carrier frequency. If a laser pulse is ultrashort, it completely covers the resonance spectrum, the population of continuous spectrum states reflects a Fano profile, and its amplitude is proportional to the spectral amplitude of a laser field at the resonance frequency. Both limiting cases allow the study of a resonance profile by its frequency scanning with a quasi-monochromatic laser field or by measuring the spectral population of continuous spectrum levels in excitation by an ultrashort pulse. The wide energy range, in which excitation of Fano resonances in various quantum structures by an electromagnetic field is observed, allows one to reckon on carrying out experiments, in which it will be possible to obtain the femtosecond resolution for the time dynamics of the asymmetric profile build-up, which may by much easier technologically than in the case of attosecond excitation of the atomic autoionization states.

**Acknowledgement**
The work was executed within the framework of the State Assignment of the RF Ministry of Education and Science (assignment No. 3.9890.2017/8.9).